\documentclass{aa}  

\usepackage{graphicx}
\usepackage{txfonts}

\usepackage{ulem}

\usepackage{xcolor}

\usepackage{soul}
\usepackage[colorlinks = true, linkcolor=blue, citecolor = blue]{hyperref}

\begin{document} 

\title{New X-ray detections of magnetic period-bounce cataclysmic variables from XMM-Newton and SRG/eROSITA}
\titlerunning{New X-ray detections of magnetic period-bounce cataclysmic variables}

   \author{Daniela Muñoz-Giraldo
          \inst{1}
          \and
          Beate Stelzer\inst{1}\textsuperscript{,}\inst{2}
          \and
          Domitilla de Martino\inst{3}
          \and
          Axel Schwope\inst{4}
          }

   \institute{Institut f\"ur Astronomie und Astrophysik, Eberhard-Karls Universit\"at T\"ubingen, Sand 1, 72076 T\"ubingen, Germany \\   
   \email{munoz-giraldo@astro.uni-tuebingen.de} 
         \and
             INAF -  Osservatorio Astronomico di Palermo, Piazza del Parlamento 1, 90134 Palermo, Italy 
          \and
          INAF - Osservatorio Astronomico di Capodimonte, Via Moiariello 16, 80131 Naples, Italy
          \and
          Leibniz-Institut für Astrophysik Potsdam (AIP), An der Sternwarte 16, 14482 Potsdam, Germany\\  
             }

   \date{Received XX; accepted XX, 2023}

  \abstract
   {A great portion of the cataclysmic variable population, between 40$\%$ and 70$\%$, is predicted to be made up of period-bouncers, systems with degenerate donors that have evolved past the period minimum.  However, due to their intrinsic faintness, only a few of these systems have been observed and confidently identified so far.}
  {We have searched for X-ray emission as a proof of accretion in order to confirm period-bounce cataclysmic variables.}
   {In this study we use data from {\it XMM-Newton} and eROSITA on a pilot sample of three period-bounce candidates with a magnetic white dwarf, which are expected to exhibit stronger X-ray emission than non-magnetic systems due to more efficient conversion of matter accretion onto the white dwarf.}
   { In a  dedicated {\it XMM-Newton} observation of the period-bounce candidate SDSS\,J151415.65+074446.5 we discovered X-ray modulation at the binary orbital period confirming it as an accreting system. The X-ray light curve and the X-ray spectrum display characteristics of magnetic Polar-type systems allowing  for the first time the determination of the X-ray luminosity and mass accretion rate for this system. Catalog data from eROSITA on the Spektrum-Roentgen-Gamma satellite for V379\,Vir and SDSS\,J125044.42+154957.4 allowed a first look into the X-ray behavior of period-bounce candidates with this new all-sky instrument. From the eROSITA measurements the X-ray luminosity and mass accretion rate were determined for the first time for SDSS\,J125044.42+154957.4, and the earlier result for V379\,Vir from {\it XMM-Newton} was confirmed.}
   {All three cataclysmic variables with a magnetic white dwarf and very low-mass donor studied in this work present evidence for X-ray emission at a similar level of $L_{\rm x}\,{\rm [erg/s]} \approx  10^{29}$, which, together with the detection of X-ray orbital modulation in two of them, V379\,Vir and SDSS\,J151415.65+074446.5, unambiguously proves the presence of accretion in these systems. The detection of these period-bouncers at faint X-ray luminosity levels with the all-sky X-ray survey eROSITA offers new prospects for the identification of additional  period-bouncers, providing impetus for theoretical studies of binary evolution.
  }


   \maketitle
%

\section{Introduction}

Cataclysmic variables (CVs) are interacting compact binaries where a white dwarf (WD) accretes matter from a Roche-lobe filling, late-type donor \citep{warner2003}. Two main types of CVs are distinguished according to the accretion geometry of the system. Usually, accretion onto the WD occurs via a disc, but if the WD magnetic field is strong enough ($\gtrsim$ 10MG), the formation of an accretion disc is prevented and accretion flow proceeds directly from the companion towards the magnetic polar regions. In these high field systems, called Polars, the WD rotation is phase-locked at the orbital period \citep{cropper}. Polars are known to switch between high and low mass accretion rate states spending roughly half their time in low-states, seen in the case of the prototype AM Her \citep{hessman2000}, making their identification and characterization challenging \citep{pretorius}.

In terms of evolution, all CVs follow a track from longer orbital periods towards shorter ones driven by angular momentum loss which causes the orbital separation of the system to decrease \citep{knigge2011}. In the course of this evolution the system passes through a period gap as a detached binary, between an orbital period ($P_{\rm orb}$) of $2-3$\,h, and re-emerges, when the donor is again filling its Roche lobe, as an active CV \citep{howell2001}. Through angular momentum loss due to gravitational radiation the evolution continues towards even shorter periods until the system reaches a period minimum located at an orbital period of $P_{\rm min} \approx 80$ \,min \citep[see e.g.][]{howell2001,patterson2011,pretorius,goliasch,mcallister2019,pala2020,pala2022}. At the period minimum the donor is out of thermal equilibrium because its mass-loss timescale is much shorter than its thermal timescale, causing the donor to stop shrinking in response to mass-loss \citep{king1988}. This results in the increase of the system's orbital separation and consequently the CV bouncing back to longer orbital periods. These systems are dubbed "period-bouncers" \citep{patterson1998}. The actual contribution from period-bouncers to the total CV population is highly disputed, with estimations ranging between 40$\%$ and 70$\%$, depending heavily on the formation and evolution model used as well as the assumptions made about the systems parameters \citep[see e.g.][]{schreiber2015,belloni2020,pala2020}. However, as a matter of fact, only a small number of period-bouncers and candidates have been observed and identified, mainly from the  Sloan Digital Sky Survey (SDSS) \citep{gansicke2009,littlefair2008}, with a few of them harbouring a magnetic WD possibly accreting at low rates \citep[see e.g.][]{schmidt2005,breedt}.

The very low-mass donors (late-M and L dwarfs) of period-bouncers have extremely weak or no coronal X-ray emission which is hardly detectable with present-day instrumentation \citep{audard2007,deLuca2020}. Therefore, the detection of X-ray emission is a key diagnostic of ongoing mass accretion in the system and hence it is the most promising path for the identification of period-bouncers. Another method for detecting mass accretion relies on the presence of H$\alpha$ emission. However, contrary to X-rays, this is not a reliable diagnostic of mass accretion in binaries with M or L dwarfs, since only upper limits to the mass accretion rate ($\dot{M}_{\rm acc}$) can be obtained due to the possible contribution of the donor’s chromospheric activity or its irradiation by the WD. 

Candidate period-bouncers can also be identified through an excess of emission in the infrared spectrum over the WD contribution. Such emission could be produced by a very late-type donor, one of the key characteristics of these systems. Since CVs with a magnetic WD typically show a higher ratio between X-ray flux and optical flux and, thus, higher X-ray luminosity as they do not present a disk through which accretion luminosity is dissipated, we have narrowed our search for period-bounce candidates to binaries composed by a magnetic WD with evidence for a very late-type companion. Of the few binaries where the presence of a very late-type companion has been confirmed spectroscopically, so far four have a magnetic WD, EF\,Eri and V379\,Vir (also known as SDSS\,J121209.31+013627.7), detected as an X-ray source by \cite{white1981} and \cite{burleigh2006} respectively, and the two siblings of V379\,Vir, SDSS\,J125044.42+154957.4 and SDSS\,J151415.65+074446.4 (henceforth SDSS\,1250 and SDSS\,1514), that have no X-ray emission reported in the literature.
The nature of EF\,Eri is, presently, still unclear. Its orbital period locates EF\,Eri near the $P_{min}$ for CVs, but the presence of a degenerate donor in the system is still debated. \cite{schwope2010} favored a  degenerate donor based on its estimated $J$ band brightness and the semi-empirical donor sequence of \cite{knigge2006}. However, with {\it Gaia}-DR3 the distance to EF\,Eri was revised to a much larger value, increasing the estimated absolute magnitude of the donor above the substellar limit and, thus, weakening the arguments for a degenerate donor in the system.

\cite{stelzer2017} detected X-ray orbital modulation on V379\,Vir from a deep {\it XMM-Newton} observation, proving that the system is accreting, and thus confirming it as a period-bounce system of the Polar type. Through modelling of the X-ray light curve, \cite{stelzer2017} derived an X-ray luminosity of $\sim 3 \times 10^{29}$\,erg/s that, assuming a mass of 0.8$M_\odot$ and a radius of $7 \times 10^8 \rm cm$ for the WD, yields a mass accretion rate in the system of $3.2 \times 10^{-14}\,{\rm M_\odot/yr}$. The estimated value for the wind driven mass accretion rate in this system is orders of magnitude too weak to explain the observed accretion rate. Additionally, donor stars in CVs are typically oversized as they are driven out of thermal equilibrium due to mass transfer \citep{knigge2011}, making Roche-lobe overflow the most likely accretion mechanism.

Yet, it still has to be demonstrated that V379\,Vir is a typical magnetic period-bouncer rather than a peculiar system. To this end, we study here the X-ray emission of the two systems (SDSS\,1250 and SDSS\,1514) composed of similar stellar components, namely a cool ($\sim$\,10000\,K) DA magnetic WD with a very low-mass companion (late-M or L estimated spectral type). Neither of them has been previously detected in the X-rays. We use {\it XMM-Newton} with its EPIC/pn detector, that offers the largest collecting area of existing X-ray instruments.

Most of the few confirmed detections of period-bouncers (see e.g. \cite{patterson2005}, \cite{mcallister2017}, \cite{neustroev2017}, \cite{pala2018}, \cite{schwope2021}, \cite{amantayeva2021}, \cite{kawka2021}, \cite{neustroev2022}) have occurred through detailed observations of specific sources using different instruments in the X-ray, optical and/or infrared bands. As a result, the sample of period-bouncers with complete and reliable system parameters is quite small, and may not be representative of the period-bouncers as a class. With the launch of the extended ROentgen Survey with an Imaging Telescope Array \citep[eROSITA;][]{predehl2021} onboard the {\it Spektrum-Roentgen-Gamma} mission \citep[SRG;][]{sunyaev2021} we are now able to conduct new studies of the X-ray activity of period-bouncers. Even though the high sensitivity of instruments like {\it XMM-Newton} is required to identify accreting period-bounce systems, the enormous statistical samples of faint X-ray sources that are being observed by eROSITA is expected to boost the number of new detections and new identifications among period-bounce candidates. In this article we carry out the analysis of the first detections of X-ray emission from period-bouncers based on the eROSITA All-Sky surveys, the last of which so far was completed in winter 2021. 

The pilot sample of period-bounce candidates is introduced in more detail in Sect.~\ref{sect:sample}. For SDSS\,1514 we present a dedicated {\it XMM-Newton} observation in Sect.~\ref{sect:sdss1514_xmm}. In Sect.~\ref{sect:sample_erosita} we discuss the detections of period-bounce candidates from this sample in the source catalogs of eROSITA. We give our conclusions in Sect.~\ref{sect:conclusions}.

\section{Sample of period-bounce candidates}\label{sect:sample}

Our pilot sample of short-period systems that we selected for our search of X-ray emission is drawn from \cite{breedt}: V379\,Vir, SDSS\,1250 and SDSS\,1514. These systems are composed by a magnetic WD and a presumably very low-mass donor. The WD has been previously identified as magnetic from SDSS spectra, specifically from the Zeeman splitting of the Balmer absorption lines (\citealt{schmidt2005}; \citealt{vanlandingham2005}; \citealt{kulebi2009}). This sample of candidates also shares some other characteristics, including large radial velocity variations, variable H$\alpha$ emission, a $\sim$\,10000\,K WD, and light curves taken with the Catalina Real-time Transient Survey that show no large-scale variability, eclipses or high states.

V379\,Vir is the only candidate with a near-IR spectrum where the donor's photospheric features of spectral type L8 were identified together with the contribution of cyclotron emission from the magnetic accreting WD \citep{farihi2008}. In absence of spectroscopic evidence, IR photometry can be used to estimate the donor spectral type of all three candidates: L5 or later for V379\,Vir estimated by \cite{schmidt2005} from comparing the absolute $J$ band magnitude of the system to observed absolute $J$ band magnitudes of L and T dwarfs; L3 or later for SDSS\,1514 estimated by \cite{breedt} from the near-IR excess in the spectral energy distribution (SED); and M8 for SDSS\,1250 estimated by \cite{steele2011} from comparing the absolute $JH$ band magnitude of the system to the observed absolute $JH$ band magnitudes of M, L and T dwarfs. At this spectral type the secondary of SDSS\,1250 would be slightly above the substellar limit on the donor sequence of \cite{knigge2006}. However, the  photometry used to estimate the donor spectral type was considered uncertain by \cite{steele2011}, making a spectroscopic confirmation  mandatory for this system.

Some relevant system properties are summarized in Table~\ref{tab:system_params}. We identify the candidates by their shortened SDSS name or variable star designation (col.~1). Col.~2 refers to the distance obtained from the {\it Gaia}-DR3 parallax, col.~3 is the orbital period derived by different authors using the H$\alpha$ emission line, col.~4 refers to the estimated spectral type of the donor, as described above, and col.~5 shows the estimated WD magnetic field strength.

\begin{table}
    \centering
     \caption{Relevant properties of our pilot sample of period-bounce candidates.}
    \label{tab:system_params}
    \begin{tabular}{ccccc} 
    \hline
    \noalign{\smallskip}
Name & $d^1$ & $P_{\rm orb}$ &  SpT$_{\rm comp}$ & $B_{\rm WD}$  \\
 & [pc] & [min]        &                 & [MG]\\
 \hline
 \noalign{\smallskip}
V379\,Vir  & $155 \pm 4$ & $88.4^2$ &  L5-L8 & $7^2$  \\
SDSS 1514  & $181 \pm 8$ & $88.7^3$ & L3    & $36^4$ \\ 
SDSS 1250  & $132 \pm 3$ & $86.3^3$ & M8   & $20^4$ \\ 
\hline
 \noalign{\smallskip}

\multicolumn{5}{l}{\footnotesize $^1$ distances from {\it Gaia}-DR3 parallax. $^2$ \cite{farihi2008}.}\\
\multicolumn{5}{l}{\footnotesize $^3$ \cite{breedt}. $^4$ \cite{kulebi2009}.}

\\
\end{tabular}
\end{table}

\subsection{Determination of the white dwarf mass}\label{subsect:wd_mass}

To estimate the individual WD radii and masses we fitted WD atmosphere models by \cite{koester2010} to the SDSS spectra\footnote{https://dr9.sdss.org/} of the period-bounce candidates in our pilot sample considering  values for $T_{\rm eff}$ and $\log(g)$ previously reported in the literature. Because these systems are magnetic WDs the Balmer lines cannot be used and our fits rely solely on the continuum. Since we select the appropriate model ($T_{\rm eff}$, $\log(g)$) beforehand, the fit has a single parameter, the ratio between the observed flux from the SDSS spectrum and the model surface flux. This ratio represents the dilution factor, $(d/R_*)^2$ where $d$ is the distance and $R_*$ the stellar radius. From the {\it Gaia}-DR3 distance we can, therefore, derive the WD radius. From the radius and an appropriate mass-radius relation we obtained the WD mass. 

For SDSS\,1514 and SDSS\,1250 we used the $10000$\,K DA WD spectrum with $\log(g)=8.0$ corresponding to the parameters derived by \cite{breedt} for the two stars from the same SDSS spectra. With the {\it Gaia}-DR3 distances in Table~\ref{tab:system_params} we obtain a WD radius of $7.0 \times 10^8$\,cm for SDSS\,1514 and of $7.2 \times 10^8$\,cm for SDSS\,1250, and from these values we derive the WD masses using the mass-radius relation by \cite{nauenberg} of 0.80$M_{\odot}$ and 0.77$M_{\odot}$ respectively.

For V379\,Vir we used a 11000\,K DA WD spectrum with $\log(g)$=8.0 following the WD temperature obtained by \cite{burleigh2006} for the same SDSS spectra and assuming a $\log(g)$ similar to the other candidates in our pilot sample. With the {\it Gaia}-DR3 distance in Table~\ref{tab:system_params} we obtain a WD radius of $8.3 \times 10^8$\,cm and from this value we derive a WD mass using the mass-radius relation by \cite{nauenberg} of 0.64$M_{\odot}$.

We report here that \cite{gentile2021} derived a WD mass for all three systems in our sample using Gaia photometry
and astrometry to fit stellar parameters. From this fitting process they obtain an effective temperature of 10668\,$\pm$\, 444\,K, 8057\,$\pm$\,1234\,K, 8211\,$\pm$\,311\,K and a $\log(g)$ of 8.0\,$\pm$\,0.1, 7.6\,$\pm$\,0.5, 7.8\,$\pm$\,0.1 for V379\,Vir, SDSS\,1514 and SDSS\,1250 respectively. Using a pure-H atmosphere model \cite{gentile2021} obtains a WD mass of 0.62\,$\pm$\,0.07$M_{\odot}$, 0.41\,$\pm$\,0.20$M_{\odot}$ and 0.48\,$\pm$\,0.05$M_{\odot}$ for V379\,Vir, SDSS\,1514 and SDSS\,1250 respectively. The values for V379\,Vir are consistent with the ones applied to the SDSS spectra. For the other two candidates, the values are considerably lower than the ones we obtained from the SDSS spectra.

\section{\textit{XMM-Newton} observation of SDSS J1514}\label{sect:sdss1514_xmm}

\textit{XMM-Newton} observed SDSS\,1514 on January 14 2020 for 41ksec (Obs-ID 0840380201; PI Stelzer) with all EPIC instruments  (\citealt{struder}, \citealt{turner}) using the \texttt{THIN} filter and with the Optical Monitor { \citep[OM;][]{mason}} in \texttt{FAST MODE} using the \textit{V} band filter.

\subsection{X-ray data}\label{subsect:xmm_xray}
 
SDSS\,1514 is detected in all EPIC instruments, giving a net count rate of 0.0042 $\pm$ 0.0004 cts/s in the MOS1 and of 0.0121 $\pm$ 0.0011\,cts/s in the MOS2 cameras, respectively. Considering how faint the source is in the MOS instrument, we limit the analysis to EPIC/pn, which provides the highest sensitivity of the EPIC detectors, as can be seen in the net source count rates presented in Table~\ref{countrate}. The data analysis was carried out with \textit{XMM-Newton's} Standard Science Analysis System (SAS) version 19.1.0. The observation is slightly affected by flaring particle background, therefore we retained only the events for which the count rate measured over the full detector area fulfills  \texttt{RATE} $\leq$ 0.6 cts/s, leaving an exposure time of 37 ks for the analysis. We filtered the data for pixel patterns (\texttt{PATTERN} $\leq$ 4), quality flag (\texttt{FLAG} = 0) and events channels (200 $\leq$ PI $\leq$ 12000). Source detection was performed in three energy bands: 0.2 - 1.0 keV (S), 1.0 - 2.0 keV (M) and 2.0 - 12.0 keV (H) using a customized procedure based on the steps implemented in the SAS task \texttt{EDETECT\textunderscore CHAIN}. 

\begin{table}
\caption{X-ray count rate and pulsed fraction (PF) from a sine fit for SDSS 1514 EPIC/pn data in different energy bands.}    
\label{countrate}      
\centering                        
\begin{tabular}{c c c c}       
\hline              
\noalign{\smallskip}
Energy     & Band & net source rate & PF\textsubscript{sine}\\ 

\mbox{[keV]}   &  label & [cts/s] &  \\
\hline                       
\noalign{\smallskip}
  0.2-12.0     & B  & 0.0335 $\pm$ 0.0013 & 1.0    \\
  0.2-1.0     & S  & 0.0181 $\pm$ 0.0009 & 1.0   \\
  1.0-2.0     & M  & 0.0096 $\pm$ 0.0006 & 1.0  \\
  2.0-12.0     & H  & 0.0058 $\pm$ 0.0006 & 0.96 $\pm$  0.02       \\
\hline                                  
\end{tabular}
\end{table}

 
For the spectral and temporal analysis we defined a circular photon extraction region with radius of $30^{\prime\prime}$ centered on the EPIC/pn source position. To ensure a homogeneous signal, the background was taken as the average of three adjacent circular regions on the same CCD chip, each with a radius of $30^{\prime\prime}$. The background substraction of the light curve was carried out with SAS task \texttt{EPICLCCORR}, which also corrects for instrumental effects, on an event list that was previously barycenter corrected using the SAS tool \texttt{BARYCEN}. 

\begin{figure}
   \centering
   \includegraphics[width=\columnwidth,trim=0.6cm 2.0cm 2.0cm 2.2cm,clip]{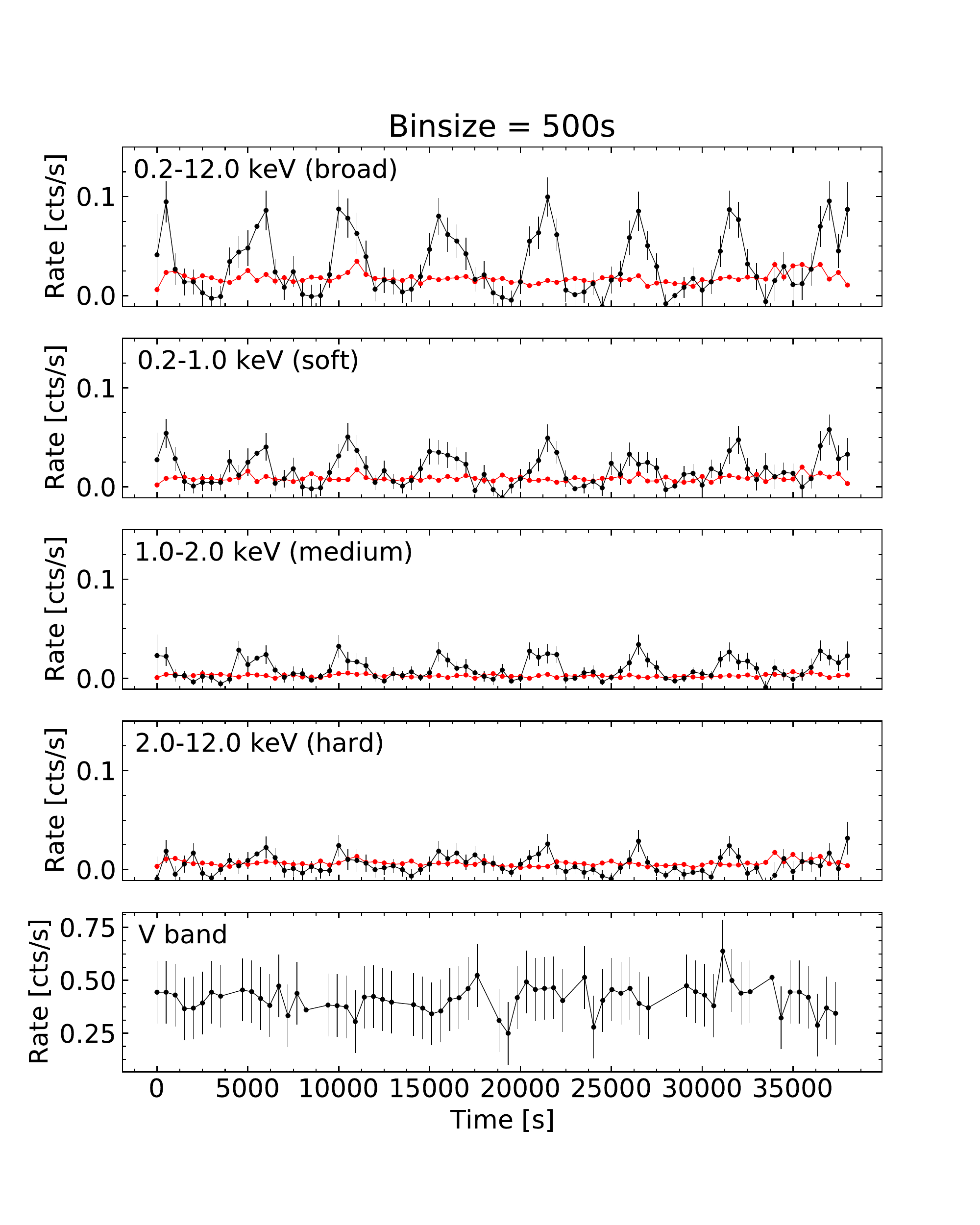}
      \caption{EPIC/pn X-ray light curve of SDSS 1514 in four energy bands, labelled in the upper left corner, and OM $V$-band light curve. The bin size is 500\,s in all panels. X-ray light curves represent the background-substracted source signal (black) and for comparison the background signal (red). 
              }
         \label{LC_Mean}
   \end{figure}

The X-ray light curve shows a clear periodic modulation in all energy bands with larger amplitude for softer emission, as can be observed in Fig.~\ref{LC_Mean}. The modulation displays an on-off behaviour that is typical for Polars \citep{cropper}. In the minimum, the count rate drops to approximately zero suggesting that the area of accretion, presumably the magnetic pole cap of the WD, is completely occulted. A Lomb-Scargle periodogram analysis of the broad band light curve yields a period of $P_{\rm orb} = 87.93 \pm  0.30$ min. This value and its 1$\sigma$ error were derived with a bootstrap approach from 5000 simulated broad band light curves that were generated by drawing the count rates of individual bins randomly from the range defined by the count rate errors. This period is in good agreement with the published period derived from the H$\alpha$ emission \citep[88.7 min;][]{breedt}. 

\begin{figure}
   \centering
   \includegraphics[width=\columnwidth,trim=0.5cm 0.6cm 2.0cm 1.5cm,clip]{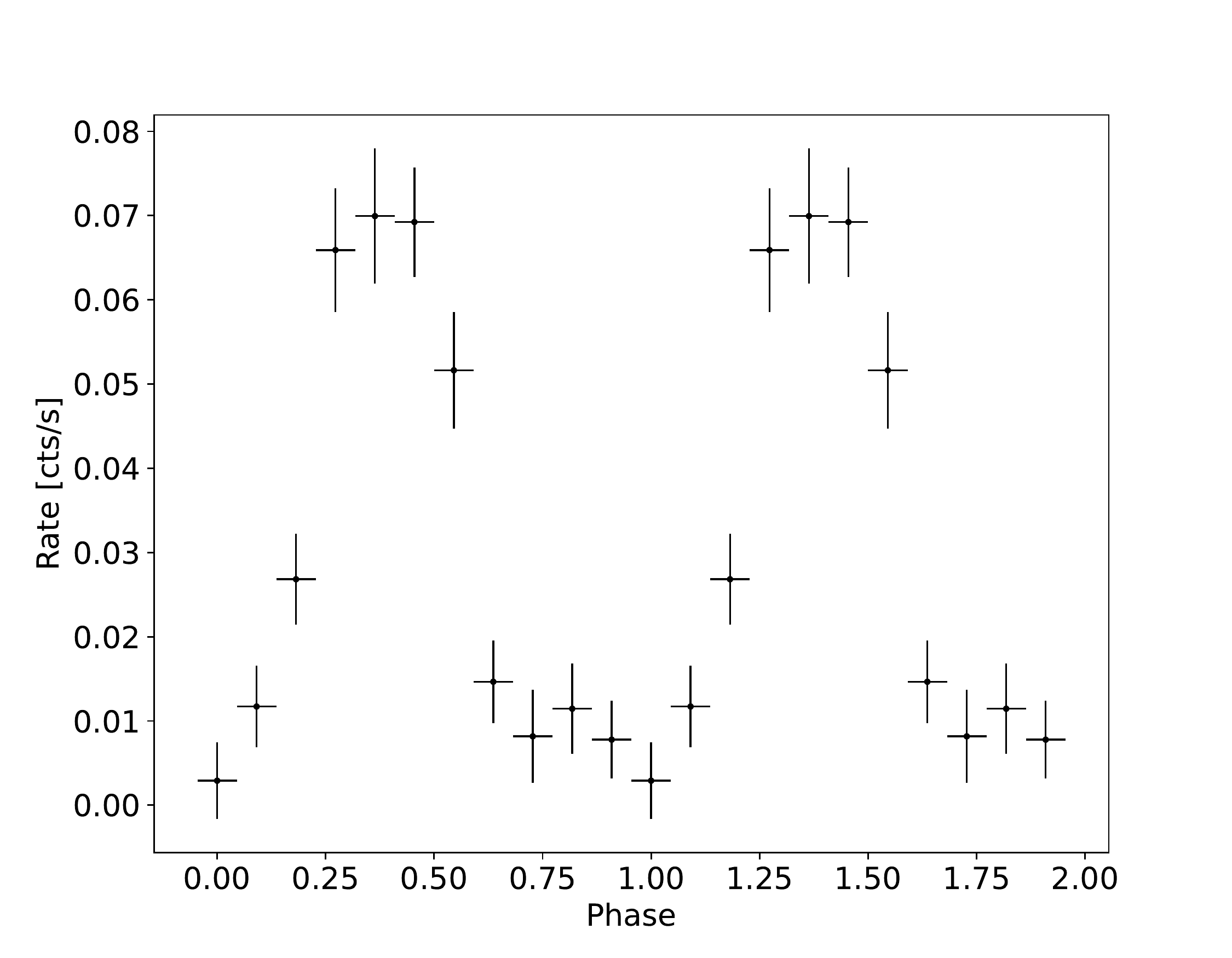}
   \\
      \caption{X-ray light curve (EPIC/pn 0.2 - 12 keV) of SDSS 1514 folded using the period determined from the X-ray signal. The bin size is 479.61s.}
         \label{PhaseError}
   \end{figure}

To first approximation, we fitted a sinusoid to the X-ray light curve from which we determined the pulsed fraction (PF) taking into account the uncertainties on y-offset and amplitude of the sine curve. The values obtained for the individual energy bands (see Table~\ref{countrate}) are consistent with the 100$\%$ PF typically observed in Polars \citep{norton1989}. The phase-folded light curve shown in Fig.~\ref{PhaseError} was obtained using the HEASARC tool \texttt{EFOLD} which uses an input of 11 phase bins to calculate a new bin time of 479.61\,s corresponding to the duration of a phase bin in the folded light curve. The folded light curve displays more clearly the already mentioned on-off behaviour with a faint phase that is slightly longer than the bright phase. 

Radial accretion at the magnetic pole can be described by the duration of the faint phase ($\gamma$) in fractions of the orbital period, where the WD is viewed at an angle $i$ to the rotation axis (the inclination), such that the magnetic pole is located at a colatitude $\beta$ (angle between the rotation and the magnetic axis) given by \cite{cropper} as 
   \begin{equation}
   \label{colatitude}
      \beta = \arctan \bigg( \frac{\cot{i} }{\cos{\pi \gamma}} \bigg)\,.
   \end{equation} 

 We do not see evidence of eclipses in the $V$ band lightcurve (Fig.~\ref{LC_Mean}), such that we can set a rough upper limit to the inclination of $i < 75^{\circ}$. Considering that we observed occultation in the X-ray light curve, a lower limit was assumed at $i > 10^{\circ}$ such that the hot-spot comes in and out of view due to the rotation of the WD. Together with the observed duration of the faint phase ($\gamma = 0.46$) determined from the folded light curve in Fig.~\ref{PhaseError} we find for the colatitude of the magnetic pole a range of $65^{\circ} < \beta < 89^{\circ}$ using Eq. (\ref{colatitude}). This result gives us information about the geometry of the system. For $\beta < 90^{\circ}$ the accreting pole is on the same side of the orbital plane as our line-of-sight. The high value for $\beta$ suggests a large misalignment between the rotation axis of the WD and its magnetic field. From Fig.~\ref{PhaseError} the shape of the folded X-ray light curve indicates that the rise and decline to the maximum are approximately of the same duration. This suggests that the emission region is symmetrical however with the available data we cannot discern if the extent is vertical or lateral.

To constrain the properties of the X-ray emitting plasma, we fitted the EPIC/pn spectrum of SDSS\,1514 using the X-Ray Spectral Fitting Package \citep[XSPEC;][]{arnaud1996}. We performed two separate fits considering, in both cases, a simple absorber ({\sc tbabs}), the first with a single-component thermal model ({\sc apec}) and the second with a thermal Bremsstrahlung model ({\sc brems}). The {\sc apec} fit was carried out using the abundances from \cite{asplund}. 

\begin{figure}
   \centering

   \includegraphics[width=\columnwidth,trim=0.7cm 1.7cm 3.3cm 2.2cm,clip]{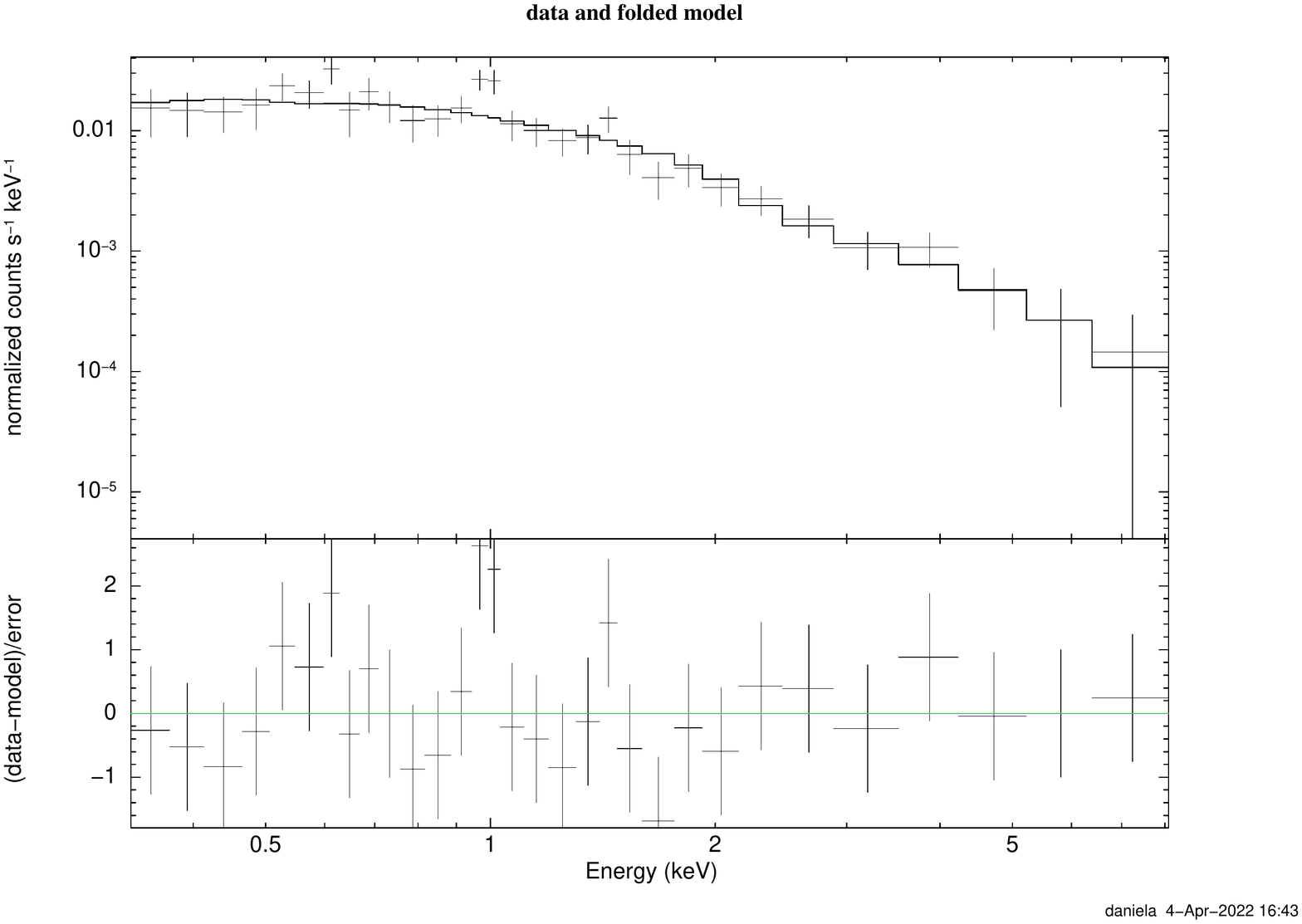}
      \caption{Time-averaged EPIC/pn X-ray spectrum of SDSS 1514 with the {\sc apec} model and residuals.
              }
         \label{Spectrum}
   \end{figure}

The {\sc brems} and the {\sc apec} models provide consistent values for the X-ray temperature of the system (see Table \ref{parameters}). For the {\sc apec} model we obtain a sub-solar abundance that is however ill-constrained, and its uncertainty also allows for zero metallicity. This might be explained by the lack of strong emission lines at the plasma temperature of about $4$\,keV as well as the low S/N ratio of the spectrum, thus justifying the use of a pure Bremsstrahlung model. The neutral hydrogen  column density also appears to be unconstrained and compatible with no absorption of the source. The range obtained from our spectral fits for the gas column density falls within the upper limit for the galactic absorption\footnote{https://heasarc.gsfc.nasa.gov/cgi-bin/Tools/w3nh/w3nh.pl} in the direction of SDSS\,1514, estimated at $2.2 \times 10^{20}$\,$\rm cm^{-2}$ \citep{bekhti2016}. The {\sc brems} and the {\sc apec} models yield similar $\chi_{\rm red}^2$, and we can not distinguish between these two models at the statistics of our observation. We show in Fig.~\ref{Spectrum} the observed EPIC/pn spectrum together with the {\sc apec} model and residuals.

\begin{table*}
\caption{Best-fit parameters for the {\it XMM-Newton} EPIC/pn spectrum of SDSS 1514 and values corresponding to upper and lower 90\% confidence ranges. The flux is given for the {\it XMM-Newton} broad band (0.2 -12 keV)}        
\label{parameters}      
\centering                         
\begin{tabular}{c c c c c c c}        
\hline               
\noalign{\smallskip}
 Model  & $\chi_{red}^2 $ (dof)   & $N_{\rm H}$ & \textit{k}T & Z & Flux & Emission measure\\ 
  &  & $[\rm cm^{-2}]$ & [keV] & [$Z_\odot$] & $[\rm erg \, \rm cm^{-2} \rm s^{-1}]$ &  $[\rm cm^{-3}]$ \\
\noalign{\smallskip}
\hline                        
\noalign{\smallskip}
 TBABS*APEC  &1.03 (27)    & $1.9_{0.0}^{6.1} \times 10^{20} $ & $3.75_{2.55}^{5.92}$ & $0.04 _{0.0}^{0.71} $& $7.20_{6.22}^{8.41} \times 10^{-14}$  & $2.16_{1.72}^{2.46} \times 10^{51} $ \\
  \noalign{\smallskip}
 TBABS*BREMSS &  1.00 (28)   & $1.6_{0.0}^{6.6} \times 10^{20} $ & $3.84_{2.38}^{6.36}$ & -- & $7.21_{6.18}^{8.37} \times 10^{-14}$ & $2.16_{1.87}^{2.66} \times 10^{51} $      \\
 \noalign{\smallskip}

\hline                                 
\end{tabular}
\end{table*}


We have also performed several {\sc apec} fits with abundances fixed at different values and found that this has a negligible influence on the X-ray flux. Therefore, we adopt an unabsorbed flux of $f_{\rm x} = 7.3\times10^{-14}$ erg/$\mbox{cm}^2$/s in the $0.2 - 12$\,keV band, taken as the average flux from all the different {\sc apec} fits we performed. For the $0.1 - 2.4$\,keV {\it ROSAT} band the average X-ray flux is $4.5 \times 10^{-14}$ erg/$\mbox{cm}^2$/s. This latter value is in accordance with the upper limit placed on the X-ray flux by the {\it ROSAT} non-detection at $\leq 6 \times 10^{-14}$ erg/$\mbox{cm}^2$/s for  this energy band \citep{breedt}. Using the best-fit {\sc apec} parameters and a dummy response which covers a broad energy range we calculated the bolometric X-ray flux to obtain a bolometric correction factor of 1.10.

With its {\it Gaia}-DR3 distance we determine the bolometric X-ray luminosity to $\log({L_{\rm x}})$\,[erg/s] $= 29.5$, and from this value we derive a mass accretion rate of ${\dot{M}}_{acc} = 3.3 \times 10^{-14}\,M_\odot$\,/\,yr obtained adopting the WD mass and radius derived in Sect.~\ref{subsect:wd_mass} for the SDSS spectra of SDSS\,1514. A mass accretion rate of ${\dot{M}}_{acc} = 9.4 \times 10^{-14}\,M_\odot$\,/\,yr was obtained adopting the WD mass and radius SDSS\,1514 from \cite{gentile2021}. Similar to the conclusion reached by \cite{stelzer2017}, the estimated lower limit for the accretion rate in this system is orders of magnitude too high to be accounted for by a wind driven mass accretion rate mechanism, making Roche-lobe overflow the most likely accretion mechanism for SDSS\,1514.

Because we are working with magnetic targets a contribution of cyclotron emission to the overall flux (and therefore luminosity) of the systems might be present. Cyclotron flux depends on several parameters like the electron temperature $kT_e$ and density $N_e$, the magnetic field strength and the angle between the line of sight and the magnetic field vector (see \cite{schwope1990}). Hence even a rough estimate of cyclotron flux needs assumptions on several unknown parameters. In addition at low accretion rates the plasma is likely in a non-hydrodynamic regime. Thus such quantification is out of scope of the present work. We, instead, prefer to report accretion rates for the sources obtained from X-ray data that should be taken henceforth as lower limits.

The EPIC/pn X-ray count rate in the minimum of the light curve (0.64\,$\leq\,\phi\,\leq$\,1.09) is 0.0083 cts/s with a standard deviation of 0.0091 cts/s, that is the system is undetected. We produced an EPIC/pn sensitivity map for the combined time intervals corresponding to the range of phases $\phi$ that define the minimum. Then we multiplied the value of the count rate of this map at the source position with the rate-to-flux conversion factor, $CF$, derived from the time-averaged X-ray spectrum. This way we obtain the upper limit to the X-ray emission at the orbital minimum of $\log{L_{\rm x,min}}$ [erg/s] $< 28.3$. The coronal emission of the only two L dwarfs detected in X-rays so far is $\log{L_{\rm x,min}}$ [erg/s] $< 27.0$ \citep{audard2007,deLuca2020}, that is we do not obtain any constraints on the L dwarf's  X-ray emission from this observation.

\subsection{Optical data}\label{subsect:xmm_optical}
 
$V$ band photometry acquired with the OM in {\sc fast mode}, simultaneously with the X-ray observation, was extracted with the SAS task \texttt{OMFCHAIN}, and the time series was barycentre corrected.
 
A Lomb-Scargle periodogram analysis of the $V$ band did not yield a significant periodicity, and when folded on the X-ray period, no phase-related variability was observed. SDSS\,1514 is optically faint, and at the time of the observation, the system was at $V_{\rm OM}$ = 18.92 $\pm$ 0.32 mag, about one magnitude above the sensitivity limit of the OM. The variability for the $V$ band was obtained by converting the OM count rates into fluxes using the conversion factor given in the {\it XMM-Newton} SAS User Guide\footnote{https://heasarc.gsfc.nasa.gov/docs/xmm/sas/USG/ommag.html}. We used the values for the average flux (${flux}_{\rm avg}$) and for the maximum flux (${flux}_{\rm max}$) in the light curve to estimate the magnitude change as  $\Delta mag = -2.5 \log_{10}({flux}_{\rm max}/{flux}_{\rm avg})$. This gives a variability amplitude of $\Delta mag  = 0.47$ mag or around $2.5\,\%$.  This percentage is consistent with the  variability estimate made by \cite{burleigh2006} for the $V$ band light curve of V379\,Vir of about 3$\%$. However, as can be seen in Fig.~\ref{LC_Mean} the variation in the OM light curve of SDSS\,1514 is not significant, due to the low statistics of the OM data.

\section{eROSITA detections}\label{sect:sample_erosita}

eROSITA has carried out four full sky surveys, called eRASS\,1 to eRASS\,4. Source catalogs from eRASS data are produced at Max Planck Institut für extraterrestrische Physik (MPE) in Garching with the eROSITA software eSASS  described by \cite{brunner2022}. These catalogs comprise all eRASS sources in the western half of the sky in terms of Galactic coordinates, that is galactic longitude $l \geq 180^\circ$, which is the sky area with german data rights. 

To obtain the highest sensitivity for detecting the presumably faint period-bounce candidates we used the merged catalog eRASS:3, that was generated from summing data from the first three all-sky surveys. The latest version of the eRASS:3 catalog available to us in December 2022 was produced  with the data processing version 020\footnote{The source catalog used in our work is\\   all{\textunderscore}s3{\textunderscore}SourceCat1B{\textunderscore}221007{\textunderscore}poscorr{\textunderscore}mpe{\textunderscore}clean.fits (for eRASS:3).}. Source detection was performed in this catalog for a single eROSITA energy band, $0.2-2.3$\,keV. 

In order to search for the period-bouncers from our pilot sample in eRASS data we performed an initial match between our targets and the {\it Gaia}-DR3 catalog. Using the corresponding {\it Gaia}-DR3 proper motions, we corrected the coordinates of the objects to the mean observing date of eRASS:3. Then we matched them with the eRASS:3 catalog allowing for a maximum separation of $30^{\prime\prime}$. After having established the closest matches for our pilot sample, we carried out a visual inspection in a $30^{\prime\prime}$ radius region around the X-ray source to assure that there where no other potential optical counterparts. In Table~\ref{eROSITAcat} we report the X-ray parameters obtained from the merged eRASS:3 catalog for the matches of the two period-bouncers with eROSITA detections, V379\,Vir and SDSS\,1250. SDSS\,1514 is located in the sky area with  exclusive Russian eROSITA data rights and, therefore, not included in the eRASS:3 catalog.
As can be seen from Table~\ref{eROSITAcat}, the sources are faint in the eRASS surveys, with only tens of counts in the merged data from the first three eROSITA surveys, yet significantly above the detection likelihood limit of $5$ set for the eRASS:3 merged catalog.

\begin{table*}
\caption{X-ray parameters from the eROSITA merged catalog eRASS:3 for V379\,Vir and SDSS\,1250 given for the eROSITA single band (0.2-2.3\,keV).}             
\label{eROSITAcat}     
\centering                         
\begin{tabular}{c c c c c c c c}        
\hline              
\noalign{\smallskip}
Name & Detection & Separation & Counts & Count rate  &  Flux ({\sc apec}) & X-ray luminosity$^*$ & Mass accretion rate$^*$\\ 

 & likelihood & [$^{\prime\prime}$] &[cts] &[cts/s] & [\rm erg $cm^{-2}$ \rm $s^{-1}$] & [erg$ s^{-1}$] & [$M_\odot \rm yr^{-1}$ ]\\
\hline                      
\noalign{\smallskip}
V379\,Vir& 44.5  & 2.2 & 30.1 $\pm$ 6.3  &  0.08 $\pm$ 0.02 &(0.75 $\pm$ 0.13) $\times 10^{-13}$ & log($\rm L_{bol}$)= 29.5 & $5.3\times 10^{-14}$   \\
  
\noalign{\smallskip}
SDSS\,1250& 19.9  & 2.9 & 16.5 $\pm$ 4.9  &  0.05 $\pm$ 0.01 &(0.46 $\pm$ 0.12) $\times 10^{-13}$ & log($\rm L_{bol}$)= 29.2 & $1.7\times 10^{-14}$   \\

\hline                                  
\noalign{\smallskip}
\multicolumn{8}{l}{\footnotesize $^*$ Values do not consider potential contributions from cyclotron emission to the X-ray luminosity}.

\end{tabular}
\end{table*}


In the eROSITA eRASS:3 catalog the single band flux has been calculated for a power law model with an index $\Gamma = 2.0 $ and a galactic absorption of $N_{\rm H} = 3 \times 10^{20} \rm cm^{-2}$ \citep{brunner2022}. Since this spectral model is not appropriate for CVs, we computed fluxes for an {\sc apec} model with help of a simulation that provided us with the eROSITA conversion factor, $CF_{\rm eROSITA,APEC}$, from count rate to thermal flux.

To this end, we generated a "fake" spectrum using the \texttt{FAKEIT} command available in XSPEC  together with the eROSITA response files. We chose an {\sc apec} model with $kT = 2.62\,$keV, $N_{\rm H} = 2.3 \times 10^{20}\, \rm cm^{-2}$ and an abundance of $0.11\, Z_\odot $, the values found from the {\it XMM-Newton} spectrum for one of our targets  \citep[see][]{stelzer2017}. We then produced another simulated eROSITA spectrum for the power law model used in the eRASS catalogs. From both synthetic spectra we then retrieved the flux and the count rate in the single band, using these values to calculate a $CF_{\rm fake,APEC}$ and a $CF_{\rm fake,powerlaw}$. The  results are shown in Table~\ref{fakeitmodel}, from where it can be seen that these two $CF$ differ by only $3.36$\,\%. This result allows us to derive $CF_{\rm eROSITA,APEC}$ from $CF_{\rm eROSITA,powerlaw}$ used in the eRASS catalogs. The final values for the {\sc apec} fluxes for the eROSITA single band (0.2-2.3 keV), given in Table~\ref{eROSITAcat}, were calculated using the catalog count rate and the {\sc apec} eROSITA conversion factor as, $Flux_{ \rm APEC}$ = Count Rate/$CF_{\rm eROSITA,APEC}$. Analogous to the analysis of the {\it XMM-Newton} data, the {\sc apec} parameters and a dummy response which covers a broad energy range were used to calculate the bolometric X-ray flux. We obtain a bolometric correction factor of 1.60. With the {\it Gaia}-DR3 distances we obtained the bolometric X-ray luminosity and mass accretion rate of each of our targets (see Table~\ref{eROSITAcat}), where we used the WD mass and radius obtained for each object in Sect.~\ref{subsect:wd_mass}. Using, instead, the values for the WD masses obtained by \cite{gentile2021} from {\it Gaia} photometry, combined with a WD radius from the \cite{nauenberg} mass-radius relation, we obtain a mass accretion rate of ${\dot{M}}_{acc} = 5.5 \times 10^{-14}\,M_\odot$\,/\,yr for V379\,Vir and of ${\dot{M}}_{acc} = 3.7 \times 10^{-14}\,M_\odot$\,/\,yr for SDSS\,1250. In the latter case, the mass accretion rate is more than twice as high than the value we calculated based on the WD mass from the SDSS spectra.

\begin{table}
    \caption{Rate-to-flux conversion factor, $CF$, in the eROSITA single band (0.2-2.3 keV) for the  simulated spectra obtained using an {\sc apec} model with $kT = 2.62\,$keV and a power law model with $\Gamma = 2.0 $. }        
    \label{fakeitmodel}     
    \centering                         
    \begin{tabular}{c c}       
    \hline              
    \noalign{\smallskip}
    $CF_{fake,APEC}$ & $CF_{fake,powerlaw}$ \\ 

    \mbox{[cts $\rm cm^{2}$ $\rm erg^{-1}$]}  & \mbox{[cts $\rm cm^{2}$ $\rm erg^{-1}$]}\\
    \hline                       
    \noalign{\smallskip}
    $4.23 \times 10^{11} $ &  $4.09 \times 10^{11} $  \\
    \noalign{\smallskip}
    \hline                                  
    \end{tabular}
\end{table}


\section{Discussion and conclusions}\label{sect:conclusions}

We have detected X-ray emission from all three magnetic period-bouncer candidates, all emitting at a similar level of $L_{\rm x} \approx  10^{29}$\,[erg/s], and hence displaying similar mass accretion rates under the assumption that the WD mass is similar in all systems. 

In \cite{stelzer2017} an {\sc apec} flux value was reported for V379\,Vir in the {\it XMM-Newton} broad band (0.2 - 12.0 keV) of 1.3 $\times 10^{-13}$ erg/$\rm cm^2$/s. Since the X-ray spectra of V379\,Vir and SDSS\,1514 are very similar we can use the {\it XMM-Newton} spectrum of SDSS\,1514 analysed in this paper to estimate the flux of V379\,Vir in the eROSITA band at its {\it XMM-Newton} epoch. In XSPEC we calculated for SDSS\,1514 a flux ratio of $1.76$  between the {\it XMM-Newton} broad band and the 0.2-2.3\,keV band used in the eRASS:3 catalog. From this we infer  the flux of V379\,Vir in the eRASS:3 energy band at the {\it XMM-Newton} epoch to be $7.1 \times 10^{-14}$ erg/$\rm cm^2$/s. A comparison with the observed eROSITA flux presented in Table~\ref{eROSITAcat} for V379\,Vir shows that there is only a $5.7\,\%$ difference between the two values and, considering the error bars, the two brightness  measurements are consistent with each other, that is the system's X-ray luminosity  (averaged over the orbital cycle) did not change on timescales of few years.

For two of the systems,V379\,Vir and SDSS\,1514, we have found X-ray orbital modulation from our long sensitive {\it XMM-Newton} observations, presented by \cite{stelzer2017} and in this paper respectively. The fluxes discussed in the previous paragraph, therefore, represent time-averages over several orbital cycles. In the case of eROSITA, the flux is the average over three surveys separated from each other by six months. Within each survey V379\,Vir and SDSS\,1250 were observed between 6 and 11 times for around 20\,s to 40\,s with a gap of $\approx4$\,h between these individual exposures, defined by the scanning law of eROSITA \citep[see e.g.][]{predehl2021}. Given the faintness of the sources we refrain here from a search of variability in eRASS data. 
 
The X-ray detection of our targets confidently proves that accretion is taking place. We have thus demonstrated that X-ray emission is an efficient way of identifying CVs with very low mass donors. While the periodic X-ray variability testifying the accretion geometry requires the high sensitivity that can be achieved only with dedicated X-ray pointings we have shown that the flux limit of the merged data from the eROSITA surveys is sufficient to provide a weak detection for systems at distances out to $\sim$\,200\,pc. A rough projection of the prospects for a systematic search for period-bouncers in the eRASS:3 catalog can be done using the X-ray flux limit of the survey and an estimate of the space-density of CVs in an exponential disk. 

To this end, we use the fainter eRASS detection among our pilot sample, SDSS\,1250, as a reference point to estimate the distance limit for period-bouncers of eRASS:3. For a detection likelihood {\sc det\_ml} $>20$, corresponding to the value for SDSS\,1250 (see Table~\ref{eROSITAcat}), $95$\,\% of the sources in the eRASS:3 catalog have a flux higher than $2 \times 10^{-14}\,{\rm erg/cm^2/s}$. Combined with the "typical" X-ray luminosity of our pilot targets ($10^{29}\,{\rm erg/s}$) this yields an approximate distance limit of $\sim$\,200\,pc for eROSITA discoveries of accreting WDs with very-low mass donors in the merged eRASS\,1 to eRASS\,3 data base. We caution that this estimate may apply only to magnetic systems which tend to be X-ray brighter than non-magnetic CVs \citep{cropper}. On the other hand, our detection likelihood threshold is very conservative. Values of {\sc det\_ml} = 11 have been shown to define samples with only modest contamination by spurious sources \citep{wolf2021}.

\begin{figure}
   \centering
   \includegraphics[width=\columnwidth]{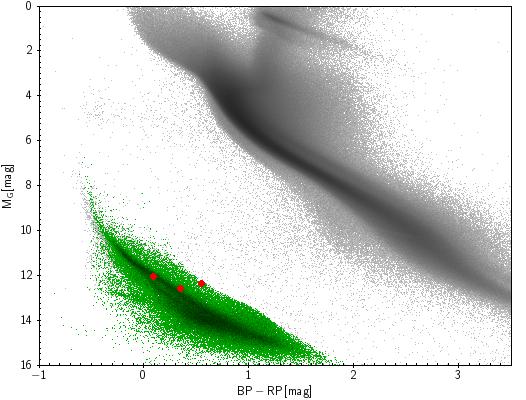}
      \caption{{\it Gaia} color-magnitude diagram showing the position of the "high fidelity" WDs from \cite{gentile2021} with a distance limit of $200$\,pc (green). The three period-bouncers from our pilot sample that are comprised in the \cite{gentile2021} catalog are shown in red. The {\it Gaia}-DR3 sources with \cite{bailer2021} distances are shown in grey as a reference.}
         \label{GentileFusillo}
\end{figure}

A possible starting point in the search of new period-bounce candidates is the {\it Gaia} catalogue of WDs by \cite{gentile2021} which presents a list of $\sim$\,360000 genuine WDs constructed through a comprehensive selection based on {\it Gaia} magnitudes and colours with consideration for diverse quality flags. In Fig.~\ref{GentileFusillo} we display as the green sample the "high fidelity" WDs with distances within the eRASS:3 limit estimated above.
The "high fidelity" WDs are objects with probability of being a WD $P_{\rm WD} \geq 0.75$ as derived by \cite{gentile2019} from the position of the objects in the {\it Gaia} color-magnitude diagram compared to the WD density map obtained from confirmed SDSS WDs. The sample shown in Fig.~\ref{GentileFusillo} in green comprises $91089$ objects. Overlaid are the three period-bouncers from our pilot X-ray survey, all of which (marked in red) are present in the \cite{gentile2021} catalog. All three are located in the same area as the {\it Gaia} WD subsample, indicating that the {\it Gaia} WD catalog by \cite{gentile2021} might include further period-bouncers. 

In Fig.~\ref{GentileFusillo} we present in grey a large sample of {\it Gaia}-DR3 sources to serve as reference. This sample was selected considering only the sources with a distance obtained by \cite{bailer2021} and a parallax error less than 1$\%$ of the parallax value.

The expected estimated number of period-bouncers within the eRASS:3 distance limit is determined by the space density of CVs, which is highly uncertain, ranging from $2 \times 10^{-4}\, \rm pc^{-3}$ \citep{de1992} to $4.8 \times 10^{-6}\, \rm pc^{-3}$ \citep{pala2020}. Considering this, for the space density of period-bouncers in the galactic mid-plane we assume $\rho_{\rm 0} \sim 10^{-5}\,{\rm pc^{-3}}$, twice the value estimated for "normal" CVs from their observed X-ray luminosity function \citep{pretorius2012}. An exponential vertical density profile with scale height of $260$\,pc for short period systems \citep{pretorius2012} then yields $\approx 133$ period-bouncers within $200$\,pc. Since the eRASS:3 catalog comprises only half of the sky, the actual expected number of PBs hidden in this catalog is $\approx67$. The number will be lower if the majority of period-bouncers has lower X-ray luminosity than the benchmark systems we studied in this work.  However, the theoretically predicted mass accretion rates \citep[on the order of $\dot{M}_{\rm acc}$ $\sim 10^{-11}$\,${\rm M_\odot/yr}$;][]{goliasch} are much higher than expected from the observed $L_{\rm x}$ values  and this suggests that the number of X-ray detections might be substantially higher than the estimate above. 

Considering the all-sky nature of eROSITA's surveys this instrument is ideal for population studies. As we have shown here, with the detections of V379\,Vir and SDSS\,1250, the eRASS data is suitable for the search for new period-bounce candidates. Once accretion has been proven in a system through an X-ray detection, an infrared (IR) spectrum should be obtained to confirm the late spectral type of the donor and, thus, to confidently verify the system as a period-bouncer. The discovery of more period-bouncers and the observational characterization of this object class is expected to provide new impetus for theoretical studies of binary  evolution.

\begin{acknowledgements}

We thank an anonymous referee for reviewing the original manuscript and giving helpful comments and useful advice. Daniela Muñoz-Giraldo acknowledges financial support from Deutsche Forschungsgemeinschaft (DFG) under grant number STE 1068/6-1. Domitilla de Martino acknowledges financial support from ASI and INAF. This work is based on observations obtained with {\it XMM-Newton}, an ESA science mission with instruments and contributions directly funded by ESA Member States and NASA. This work is based on data from eROSITA, the primary instrument aboard SRG, a joint Russian-German science mission supported by the Russian Space Agency (Roskosmos), in the interests of the Russian Academy of Sciences represented by its Space Research Institute (IKI), and the Deutsches Zentrum f\"ur Luft- und Raumfahrt (DLR). The SRG spacecraft was built by Lavochkin Association (NPOL) and its subcontractors, and is operated by NPOL with support from the Max Planck Institute for Extraterrestrial Physics (MPE). The development and construction of the eROSITA X-ray instrument was led by MPE, with contributions from the Dr. Karl Remeis Observatory Bamberg and ECAP (FAU Erlangen-N\"urnberg), the University of Hamburg Observatory, the Leibniz Institute for Astrophysics Potsdam (AIP), and the Institute for Astronomy and Astrophysics of the University of T\"ubingen, with the support of DLR and the Max Planck Society. The Argelander Institute for Astronomy of the University of Bonn and the Ludwig Maximilians Universit\"at M\"unchen also participated in the science preparation for ero. The eROSITA data shown here were processed using the eSASS/NRTA software system developed by the German eROSITA consortium. This work has made use of data from the European Space Agency (ESA) mission {\it Gaia} (\url{https://www.cosmos.esa.int/gaia}), processed by the {\it Gaia} Data Processing and Analysis Consortium (DPAC, \url{https://www.cosmos.esa.int/web/gaia/dpac/consortium}). Funding for the DPAC has been provided by national institutions, in particular the institutions participating in the {\it Gaia} Multilateral Agreement. 

\end{acknowledgements}

%
%

\bibliography{references}
\bibliographystyle{aa} 

\end{document}